\documentclass[aps,preprint,showpacs,floatfix]{revtex4-1}
\usepackage{bm}
\usepackage{amsmath}
\usepackage{epsfig}
\usepackage{rotating}
\usepackage{graphicx}

\def\be{\begin{equation}}

\def\ee{\end{equation}}
\def\barr{\begin{array}}
\def\earr{\end{array}}

\def\l{\left}
\def\r{\right}
\def\dis{\displaystyle}
\def\ed{\end{document}}

\def\can{{\cal N}}

\def\la{\lambda}
\def\ed{\end{document}}
\begin{document}

\title{Simple formula for leading $SU(3)$ irreducible representation for 
nucleons in an oscillator shell}

\author{V.K.B. Kota\footnote{
Phone:+917926314939, Fax:+917926314460  \\ {\it E-mail address:}
vkbkota@prl.res.in (V.K.B. Kota)}}

\affiliation{Physical Research  Laboratory, Ahmedabad 380 009, India}

\begin{abstract}

Applications of rotational $SU(3)$ symmetry in nuclei, using Elliott's $SU(3)$ or
pseudo-$SU(3)$ or proxy-$SU(3)$ model, often need just the lowest or leading
$SU(3)$ irreducible representation (irrep) $(\lambda_H, \mu_H)$. For nucleons in
an oscillator shell $\eta$, with $\can=(\eta +1)(\eta +2)/2$, we have the
algebra $U(r\can) \supset [U(\can \supset SU(3)] \otimes SU(r)$; $r=2$ when
there are only valence protons or neutrons and $r=4$ for nucleons with isospin
$T$. Presented in this paper is a simple general formula for the leading $SU(3)$
irrep $(\lambda_H, \mu_H)$ in any given irrep $\{f\}$ of $U(\can)$. Results are
provided for $(\lambda_H, \mu_H)$ irreps for $\eta$ values of interest in nuclei
and for this for all allowed particle numbers. These results clearly show that
prolate shape dominates over oblate shape in the shell model $SU(3)$
description. 

\end{abstract}


\maketitle

\section{Introduction}

Elliott has recognized way back in 1958 that there is $SU(3)$ symmetry
generating rotational spectra in nuclei within the spherical shell model (SM)
picture and this is found to be good for light nuclei \cite{Ell-58}.  With the
strong spin-orbit force breaking Elliott's $SU(3)$ symmetry, Draayer et al
recognized \cite{Hecht1,Hecht2,JPD1,JPD2} that for heavier nuclei pseudo-$SU(3)$,
based on pseudo spin and pseudo Nilsson orbits, will apply. More recently,
proxy-$SU(3)$ scheme by Bonatsos et al \cite{Bona-1,Bona-2,Bona-3} again brought
$SU(3)$ symmetry into focus. In all these, the lowest or the leading $SU(3)$
irreducible representation (irrep), labeled in Elliott's notation by
$(\lambda_H, \mu_H)$,  plays a significant role. For example, in light nuclei
such as $^{20}$Ne the leading irrep $(80)$ describes the ground $K^\pi=0^+$
band, in $^{24}$Mg the leading irrep $(84)$ describes the ground $0^+$ and the
$2^+$ $\gamma$-band and so on \cite{Harvey,JPD3}. Similarly, using the
pseudo-$SU(3)$ model with leading $SU(3)$ irreps for valence protons and
neutrons respectively, low-lying rotational bands are described with examples
from rare-earth and actinides nuclei \cite{JPD1}. This gave rise to integrity
basis spectroscopy \cite{JPD3,JPD-lectu}. More strikingly, leading $SU(3)$ irrep
in the proxy-$SU(3)$ scheme is used to describe prolate shape dominance over
oblate shape in nuclei with protons and neutrons occupying different oscillator
shells and in some situations the same shell \cite{Bona-2,Bona-3}. 

With nucleons in a oscillator major shell $\eta$, we have $U(r\can) \supset
[U(\can) \supset SU(3) \supset SO(3)] \otimes SU(r)$; $\can = (\eta +2)(\eta
+2)/2$ and $r=2$ for identical nucleons and 4 for nucleons with isospin. 
Computer codes for determining the $SU(3)$ irreps contained in a $U(\can)$ irrep
are available \cite{VKB1,VKB2,JPD4} and obviously they give the leading $SU(3)$
irrep besides giving all other irreps with their multiplicities. However, due to
their importance, it is clearly useful to have a formula for obtaining the leading
irrep $(\lambda_H, \mu_H)$ rather than employing detailed computer codes. The
purpose of the present note is to report a simple formula for $(\lambda_H,
\mu_H)$ and provide tabulations for useful examples with $r=2$ and $r=4$.  

In Section II formula for the leading $SU(3)$ irrep is given along with a
tabulation for identical nucleons in oscillator shells $\eta = 2-6$. An
application of these results is for nuclei with protons and neutrons in
different shells. Section III gives results for proton-neutron systems and their
applications. Finally, Section IV gives conclusions. 

\section{Formula for leading $SU(3)$ irrep: Results for identical nucleons}

Given a oscillator shell number $\eta$, we have $U((\eta +1)(\eta +2)/2) \supset
SU(3)$. In the reminder of this paper we will use $\can = (\eta +1)(\eta +2)/2$.
In nuclei we are interested in $\eta$ up to $6$ or $7$. Now, the problem we want
to solve is to find the leading (or ground state) $SU(3)$ irrep  $(\lambda_H ,
\mu_H)$ in the reduction of a irrep $\{f\}$ of $U(\can)$. Note that
$\{f\}=\{f_1,f_2,\ldots,f_\can\}$ and $f_1 \geq f_2 \geq f_3 \geq \ldots \geq
f_\can \geq 0$. For example, for $m$ identical nucleons (protons or neutrons
with spin 1/2 degree of freedom) in the oscillator shell $\eta$, $\{f\}=\{2^p\}$
for $m$ even and total spin $S=0$ with $p=m/2$. Similarly, $\{f\}= \{2^p,1\}$
for $m$ odd and $S=1/2$ with $p=(m-1)/2$; note that $m=\sum_i f_i$. For a given
$\{f\}$, the irrep $(\lambda_H , \mu_H)$ is the one with largest value of
$\epsilon=2\lambda_H + \mu_H$ and for this $\epsilon$ with largest value of
$\Lambda=\frac{\mu_H}{2}$ \cite{He-lectu}.

For a given $\eta$, the single particle orbits in $(n_z , n_x , n_y)$
representation are $(\eta -r, r-x, x)$ in the order with $r$ taking values $0$
to $\eta$ and for each $r$, $x$ takes values $0$ to $r$; note that $n_i$ gives
number of oscillator quanta in $i$-th direction for a single particle. For
example for $\eta=3$, in order they are $(300)$, $(210)$, $(201)$, $(120)$,
$(111)$, $(102)$, $(030)$, $(021)$, $(012)$ and $(003)$. Now, let us consider an
irrep $\{f\}$ for $m$ particles. Putting $f_1$ number of particles in the first
orbit, $f_2$ in the second orbit and so on with $f_\can$ in the last orbit will
give the total $N_z$, $N_x$ and $N_y$ (note that $N_i$ is the sum of single particle
$n_i$ values). Then, we have $\lambda_H=N_z-N_x$ and $\mu_H=N_x-N_y$
\cite{VKB3}. This, we are able to convert into a simple formula that is valid for
any $\eta$ and $\{f\}$. The final result is 
\be 
\barr{l} \l\{f\r\}_{U((\eta
+1)(\eta +2)/2)} \rightarrow (\lambda_H , \mu_H)_{SU(3)}  \;\;\mbox{with} \\
\lambda_H = \dis\sum_{r=0}^\eta \dis\sum_{x=0}^r (\eta -2r+x) \times 
f_{1+x+\frac{r(r+1)}{2}}\;,\;\;\;\;\mu_H=\dis\sum_{r=0}^\eta \dis\sum_{x=0}^r 
(r-2x) \times f_{1+x+\frac{r(r+1)}{2}}\;. 
\earr 
\label{su3formu} 
\ee 
It is trivial to programme Eq. (\ref{su3formu}). Results obtained using this,
for shells $ \eta = 2,3,4,5,6$ for identical fermions with particle number $m=2$
to $\can$ and spin $S=0$ for even $m$ and $S=1/2$ for odd $m$, are given in
Table I. They can be extended easily to $\eta=7,8$ as may be needed for SHE
nuclei and also for other spin $S$ values. In  \cite{Bona-2}, detailed code given
in \cite{JPD4} that generates the full $\{f\} \rightarrow  (\lambda \mu)$ reductions
is used to obtain $(\la_H , \mu_H)$ (see Table I of \cite{Bona-2}).

\section{Application to heavy nuclei with valence protons and neutrons in
different shells using proxy-$SU(3)$ scheme}

Using the results for identical nucleons given in Table I, it is possible to
infer prolate to oblate transition, in heavy nuclei with protons and neutrons in
different shells, as number of valence nucleons in a given shell changing. This
is indeed possible as  nuclei will be prolate for $\lambda > \mu$ and oblate for
$\lambda < \mu$ \cite{beta-gam1,beta-gam2,beta-gam3}. Here, we will use the
$SU(3)$ symmetry extension to heavy nuclei by the proxy-$SU(3)$ model introduced
in \cite{Bona-1,Bona-2}. Firstly, for identical nucleons in shells $\eta=2$, 3,
4, 5, 6 prolate shape changes to oblate at $m=8$, 15, 23, 34 and 47 respectively
as seen from the $(\lambda_H ,  \mu_H)$ values shown in Table I for each $\eta$.
Note that these $m$ values are higher than the mid-shell values  (6, 10, 15, 21
and 28 respectively) and therefore prolate to oblate transition breaks
particle-hole symmetry. For heavy nuclei with valence protons ($p$) and neutrons
($n$) occupying different shell say $\eta_p$ and $\eta_n$, the leading or ground
state $SU(3)$ irrep will be $(\lambda_H^p + \lambda_H^n , \mu_H^p + \mu_H^n)$
where  $(\lambda_H^p, \mu_H^p)$ and $(\lambda_H^n , \mu_H^n)$ are the leading
$SU(3)$ irreps for protons and neutrons respectively and they can obtained from
Table I. This is employed in the proxy-$SU(3)$ model and here for Ba to Pt
isotopes with $88 \leq N \leq 120$, valence protons will be in 50-82 shell
($sdg$ or $\eta=4$ in the proxy model) and neutrons in 82-126 shell ($fph$ or
$\eta=5$ in the proxy model). Then, it is easy to see that the oblate shape 
($\lambda_H > \mu_H$) appears for $^{190,192,194}$W, $^{192,194,196}$Os and
$^{194,196,198}$Pt. From experiments there are strong indications that 
$^{190}$W, $^{192,194}$Os are oblate or more towards oblate shape
\cite{Expt1,Expt2,Expt3}. Using the results in Table I, it is also possible to
predict that some of the Dy, Er  and HF isotopes (with neutron number N between
88 and 120) also exhibit oblate shape. See \cite{Bona-2,Bona-3} for  details of
prolate-oblate transition within the proxy-$SU(3)$ model. 

\section{Leading $SU(3)$ irrep for proton-neutron systems with good spin-isospin
$SU(4)$ symmetry}

With protons and neutrons in the same oscillator shell $\eta$, we need to
consider four column irreps of $U(\can)$ and find $(\lambda_H , \mu_H)$ contained
in these irreps. Firstly we have the algebra $U((4\can) \supset U(\can) \otimes
SU(4)$; here we assume that the Wigner's spin-isospin $SU(4)$ symmetry is good.
Given the nucleon number $m$ and the isospin $T=|T_Z|$ (note that
$T_Z=(m_p-m_n)/2$ where $m_p$ is number of valence protons and $m_n$ is number 
of valence
neutrons with $m=m_p+m_n$), the lowest $U(4)$ irrep $\{F_1,F_2,F_3,F_4\}$ is 
given as follows; note that $SU(4)$ irreps follow from $U(4)$ irreps.  For $m$
even we have
\be
\barr{l}
\l\{F_1,F_2,F_3,F_4\r\} = \l\{\frac{m+2T}{4}, \frac{m+2T}{4}, \frac{m-2T}{4}, 
\frac{m-2T}{4}\r\}\;\;\;\mbox{for}\;\;\;\frac{m}{2}+T\;\;\mbox{ even}\;,\\
\l\{F_1,F_2,F_3,F_4\r\}=\l\{\frac{m+2T+2}{4}, \frac{m+2T-2}{4}, 
\frac{m-2T+2}{4}, \frac{m-2T-2}{4}\r\}\;\;\;\mbox{for}\;\;\;
\frac{m}{2}+T\;\;\mbox{odd}\;.\\
\earr \label{su4-1}
\ee
The only exception is $T=0$ for $m=4r+2$ type and then 
\be
\l\{F_1,F_2,F_3,F_4
\r\}=\l\{\frac{m+2}{4}, \frac{m+2}{4}, \frac{m-2}{4}, \frac{m-2}{4}\r\}\;.
\label{su4-2}
\ee
Similarly, for odd-$m$ we have
\be
\barr{l}
\l\{F_1,F_2,F_3,F_4\r\}=\l\{\frac{m+2T+2}{4}, \frac{m+2T-2}{4}, \frac{m-2T}{4},
\frac{m-2T}{4}\r\}\;\;\;\mbox{for}\;\;\;\frac{m}{2}+T\;\;\mbox{odd}\;,\\
\l\{F_1,F_2,F_3,F_4\r\}=\l\{\frac{m+2T}{4}, \frac{m+2T}{4}, \frac{m-2T+2}{4}, 
\frac{m-2T-2}{4}\r\}\;\;\;\mbox{for}\;\;\;\frac{m}{2}+T\;\;\mbox{even}\;.\\
\earr \label{su4-3}
\ee
Eqs. (\ref{su4-1})-(\ref{su4-3}) are well known and given in many papers (but 
with different notations). See for example Refs. \cite{Manan-su4,Piet-su4}.
Using  Eqs. (\ref{su4-1})-(\ref{su4-3}) it is easy to obtain the lowest $U(4)$
irrep for a given $m$  and $|T_z|$. Note that, with
$\can^\prime=(\eta+1)(\eta+2)$, values of $T_Z$ for  $m \leq \can^\prime$  are
$|T_Z|=m/2$, $m/2-1$,\ldots, $0$ or $1/2$. For $m >\can^\prime$, we have 
$|T_Z|=(2\can^\prime -m)/2$, $(2\can^\prime -m)/2-1$,\ldots, $0$ or $1/2$.  Now,
our task is to find the $U(\can)$ irrep $\{f\}$ that corresponds to a given
$\l\{F_1,F_2,F_3,F_4\r\}$. It is well known that $\{f\}$ must be conjugate of
$\l\{F_1,F_2,F_3,F_4\r\}$ and it can be identified as follows. Firstly, $\{f\}$
will be $\{4^a 3^b 2^c 1^d\}$ type irrep. Then, it is easy to find that $a=F_4$,
$b=F_3-F_4$, $c=F_2-F_3$ and $d=F_1-F_2$. Therefore, $\{f\}$ will have, in Young
tableaux notation, four boxes in $a$ ($=F_4$) number of rows, 3 boxes in $b$
($=F_3-F_4$) number of rows, two boxes in $c$ ($=F_2-F_3$) number of rows and
one box in $d$ ($=F_1-F_2$) number of rows. With $\{f\}$ thus determined, we can
find the leading $SU(3)$ irrep in the four columned $\{f\}$ using Eq.
(\ref{su3formu}). Results are given in Tables II-IV for $\eta=2$, $3$ and $4$.
These results will suffice as they are needed only for nuclei with valence
nucleons in the same oscillator shell. Let us now consider some applications of
these results.

Firstly, for N=Z and N=Z $\pm$ 1 nuclei, it is easy to see that the leading or
ground state $U(\can)$ irrep $\{f\}$ will be of $\{4^r,p\}$ type with $\{4^r\}$
for $m=4r$ (i.e. N=Z even-even), $\{4^r,2\}$ for $m=4r+2$ (N=Z odd-odd) and
$\{4^r,1\}$ or $\{4^r,3\}$ as appropriate for $m=4r+1$ or $4r+3$ (N=Z $\pm$ 1).
The leading $SU(3)$ irrep for these for $\eta=2$, 3 and 4 shells are listed in
Table II and they will suffice in practical applications. It is seen that the
prolate to oblate transition here is for $m=16$, $29$ and $46$ for $\eta=2$, $3$
and $4$ shells respectively. Let us consider $^{72}$Kr as an example. For this
nucleus, ground state is oblate if we consider the valence nucleons to be only
in $^2p_{3/2}$, $^1f_{5/2}$ and $^2p_{1/2}$ orbits with $^{56}$Ni core (then
$\eta=2$ will apply with pseudo $SU(3)$). However, if use the proxy-$SU(3)$ model and include the
$^1g_{9/2}$ orbit (without the $\Omega=\pm 9/2$ states), then $\eta=3$ shell
will apply giving prolate shape. Experimental data for this nucleus indicates
prolate with some oblate mixing; see for example \cite{KS,Kr72}. Let us also add
that heavy N=Z nuclei (more so odd-odd N=Z nuclei) may not preserve $SU(3)$
symmetry and here $SO(N)$ structures from IBM-3 and IBM-4 models may be more
appropriate \cite{IBM-3,IBM-4}.  Turning to N $\neq$ Z nuclei but with valence
protons and neutrons in the same shell, the leading $SU(3)$ irrep for these
follow from Tables III and IV and their application is discussed in
\cite{Bona-2} with examples from Ba (Z=56) to Os (Z=78) isotopes. Results in
Table III of \cite{Bona-2} follow from Tables III and IV in the present paper.
We will discuss in a future publication in detail the dominance of prolate over
oblate shape that can be inferred from Tables I-IV and the experimental evidence for
these. 

\section{Conclusions}

Following the significance of leading $SU(3)$ irrep in Elliott's $SU(3)$,
pseudo-$SU(3)$ and proxy-$SU(3)$ models, we have presented a simple formula as
given by Eq. (\ref{su3formu}) for the $(\lambda_H , \mu_H)$ for nucleons in a
given oscillator shell $\eta$. Tabulations (Tables I-IV) are provided for values of $\eta$
that are important for nuclei. They will cover all the situations with (i)
valence protons and neutrons occupying different shells and (ii) valence protons
and neutrons occupying the same shell as in N $\sim$ Z nuclei and in many
other situations as given in Table III of \cite{Bona-2}. Eq. (\ref{su3formu})
eliminates the need to use detailed computer codes for obtaining $(\la_H ,
\mu_H)$. A brief discussion,
based on the results in Tables I-IV, of prolate dominance over oblate shape is
given with detailed discussion postponed to a separate paper.

It is useful to add that Eq. (\ref{su3formu}) is also useful in IBM-2, IBM-3 and
IBM-4 models \cite{Iac-87,KS}. In IBM-2 with $F$-spin we have $U(2\can) \supset
[U(\can) \supset SU(3)] \otimes SU(2)$ and therefore $\{f\}$ of $U(\can)$ with
two rows are allowed. In IBM-3 with $U(3\can) \supset [U(\can) \supset SU(3)]
\otimes  SU_T(3)$, $\{f\}$ of $U(\can)$ here can have maximum three rows (the
other $SU(3)$ here generates isospin  of the bosons). Similarly, in IBM-4 with
$U(6\can) \supset  [U(\can) \supset SU(3)] \otimes SU(6)$ we have $\{f\}$ of
$U(\can)$ with  maximum six rows. Note that $\can=6$ for $sd$IBM, $15$ for
$sdg$IBM and so on. Applications of  Eq. (\ref{su3formu}) for these will be
discussed elsewhere. 

\acknowledgments

Thanks are due to D. Bonatsos for useful correspondence. Motivation for the work
came from the talks in the Sofia meetings held in the years 2015, 2016 and 2017
and thanks are due to N. Minkov for these meetings.

\begin{table}
\begin{center}
\caption{Ground state or leading $SU(3)$ irrep $(\lambda_H , \mu_H)$ for a given 
number $m$ of identical particles in a oscillator shell $\eta$. Results are 
shown for $\eta = 2$, 3, 4, 5 and 6. For a given $m$ with $m \geq 4$, the  
$(\lambda_H , \mu_H)$ is given in the table as $(\lambda_H , \mu_H)^m$.}
\begin{tabular}{l}
\hline
$\eta=2$ \\
$( 4, 0)^ 2$,$( 4, 1)^ 3$,$( 4, 2)^ 4$,$( 5, 1)^ 5$,$( 6, 0)^ 6$,$( 4, 2)^ 7$, 
$( 2, 4)^ 8$,$( 1, 4)^ 9$,$( 0, 4)^{10}$,$( 0, 2)^{11}$,$( 0, 0)^{12}$ \\
$\eta=3$ \\
$( 6, 0)^ 2$,$( 7, 1)^ 3$,$( 8, 2)^ 4$,$(10, 1)^ 5$,$(12, 0)^ 6$,$(11, 2)^ 7$,
$(10, 4)^ 8$,$(10, 4)^ 9$,$(10, 4)^{10}$,$(11, 2)^{11}$,\\
$(12, 0)^{12}$,$( 9, 3)^{13}$,
$( 6, 6)^{14}$,$( 4, 7)^{15}$,$( 2, 8)^{16}$,$( 1, 7)^{17}$,$( 0, 6)^{18}$,
$( 0, 3)^{19}$,$( 0, 0)^{20}$ \\
$\eta=4$ \\
$( 8, 0)^ 2$,$(10, 1)^ 3$,$(12, 2)^ 4$,$(15, 1)^ 5$,$(18, 0)^ 6$,$(18, 2)^ 7$,
$(18, 4)^ 8$,$(19, 4)^ 9$,$(20, 4)^{10}$,$(22, 2)^{11}$,$(24, 0)^{12}$,\\
$(22, 3)^{13}$,
$(20, 6)^{14}$,$(19, 7)^{15}$,$(18, 8)^{16}$,$(18, 7)^{17}$,$(18, 6)^{18}$,
$(19, 3)^{19}$,$(20, 0)^{20}$,$(16, 4)^{21}$,$(12, 8)^{22}$,\\
$( 9,10)^{23}$,$( 6,12)^{24}$,
$( 4,12)^{25}$,$( 2,12)^{26}$,$( 1,10)^{27}$,$( 0, 8)^{28}$,$( 0, 4)^{29}$,
$( 0, 0)^{30}$\\
$\eta=5$ \\
$(10, 0)^ 2$,$(13, 1)^ 3$,$(16, 2)^ 4$,$(20, 1)^ 5$,$(24, 0)^ 6$,$(25, 2)^ 7$,
$(26, 4)^ 8$,$(28, 4)^ 9$,$(30, 4)^{10}$,$(33, 2)^{11}$,$(36, 0)^{12}$,\\
$(35, 3)^{13}$,
$(34, 6)^{14}$,$(34, 7)^{15}$,$(34, 8)^{16}$,$(35, 7)^{17}$,$(36, 6)^{18}$,
$(38, 3)^{19}$, $(40, 0)^{20}$,$(37, 4)^{21}$,$(34, 8)^{22}$,\\
$(32,10)^{23}$,$(30,12)^{24}$,$(29,12)^{25}$,
$(28,12)^{26}$,$(28,10)^{27}$,$(28, 8)^{28}$,$(29, 4)^{29}$,$(30, 0)^{30}$,
$(25, 5)^{31}$,\\
$(20,10)^{32}$,$(16,13)^{33}$,$(12,16)^{34}$,$( 9,17)^{35}$,$( 6,18)^{36}$,
$( 4,17)^{37}$,$( 2,16)^{38}$,$( 1,13)^{39}$,$( 0,10)^{40}$,\\
$( 0, 5)^{41}$,$( 0, 0)^{42}$\\
$\eta=6$ \\
$(12, 0)^ 2$,$(16, 1)^ 3$,$(20, 2)^ 4$,$(25, 1)^ 5$,$(30, 0)^ 6$,$(32, 2)^ 7$,
$(34, 4)^ 8$,$(37, 4)^ 9$,$(40, 4)^{10}$,$(44, 2)^{11}$,$(48, 0)^{12}$,\\
$(48, 3)^{13}$,
$(48, 6)^{14}$,$(49, 7)^{15}$,$(50, 8)^{16}$,$(52, 7)^{17}$,$(54, 6)^{18}$,
$(57, 3)^{19}$,$(60, 0)^{20}$,$(58, 4)^{21}$,$(56, 8)^{22}$,\\
$(55,10)^{23}$,$(54,12)^{24}$,$(54,12)^{25}$,$(54,12)^{26}$,$(55,10)^{27}$,
$(56, 8)^{28}$,$(58, 4)^{29}$,$(60, 0)^{30}$,$(56, 5)^{31}$,\\
$(52,10)^{32}$,
$(49,13)^{33}$,$(46,16)^{34}$,$(44,17)^{35}$,$(42,18)^{36}$,$(41,17)^{37}$,
$(40,16)^{38}$,$(40,13)^{39}$,$(40,10)^{40}$,\\
$(41, 5)^{41}$,$(42, 0)^{42}$,
$(36, 6)^{43}$,$(30,12)^{44}$,$(25,16)^{45}$,$(20,20)^{46}$,$(16,22)^{47}$,
$(12,24)^{48}$,$( 9,24)^{49}$,\\
$( 6,24)^{50}$,$( 4,22)^{51}$,$( 2,20)^{52}$,
$( 1,16)^{53}$,$( 0,12)^{54}$,$( 0, 6)^{55}$,
$( 0, 0)^{56}$\\
\hline
\end{tabular}
\end{center}
Note that for $m=2$ we have $\{f\}=\{2\}$ and $\{1^2\}$ and here complete
solution for $\{f\} \rightarrow \lambda \mu)$ is easy to derive. We have $\{f\}
\rightarrow (\lambda \mu)= (2m,0) \oplus (2m-4,2) \oplus (2m-8,4) \oplus \ldots,
$. Similarly, $\{1^2\} \rightarrow (2m-2,1) \oplus (2m-6,3) \oplus \ldots$.  
\end{table}
\begin{table}
\begin{center}
\caption{Leading $SU(3)$ irrep $(\lambda_H , \mu_H)$ for a given 
number $m$ of nucleons with $\{f\}=\{4^r,p\}$ irrep for $U((\eta +1)(\eta
+2)/2)$ in a oscillator shell $\eta$; $m=4r+p$. Note that uniquely: (i) for N=Z
even-even nuclei lowest $\{f\}=\{4^r\}$; (ii) for N=Z odd-odd nuclei, lowest
$\{f\}=\{4^r,2\}$; (iii) for N=Z $\pm$ 1 nuclei, lowest $\{f\}=\{4^r,1\}$ or
$\{4^r,3\}$. Results are  shown for $\eta = 2$, 3, and 4. For a given $m$, the
$(\lambda_H , \mu_H)$ is given in the table as $(\lambda_H , \mu_H)^m$.}
\begin{tabular}{l}
\hline
$\eta=2$ \\
$( 4, 0)^{ 2}$,$( 6, 0)^{ 3}$,$( 8, 0)^{ 4}$,$( 8, 1)^{ 5}$,$( 8, 2)^{ 6}$,
$( 8, 3)^{ 7}$,$( 8, 4)^{ 8}$,$( 9, 3)^{ 9}$,$(10, 2)^{10}$,\\
$(11, 1)^{11}$,$(12, 0)^{12}$,$(10, 2)^{13}$,$( 8, 4)^{14}$,$( 6, 6)^{15}$,
$( 4, 8)^{16}$,$( 3, 8)^{17}$,$( 2, 8)^{18}$,$( 1, 8)^{19}$,\\
$( 0, 8)^{20}$,$( 0, 6)^{21}$,$( 0, 4)^{22}$,$( 0, 2)^{23}$,$( 0, 0)^{24}$\\
$\eta=3$ \\
$( 6, 0)^{ 2}$,$( 9, 0)^{ 3}$,$(12, 0)^{ 4}$,$(13, 1)^{ 5}$,$(14, 2)^{ 6}$,
$(15, 3)^{ 7}$,$(16, 4)^{ 8}$,$(18, 3)^{ 9}$,$(20, 2)^{10}$,\\
$(22, 1)^{11}$,$(24, 0)^{12}$,$(23, 2)^{13}$,$(22, 4)^{14}$,$(21, 6)^{15}$,
$(20, 8)^{16}$,$(20, 8)^{17}$,$(20, 8)^{18}$,$(20, 8)^{19}$,\\
$(20, 8)^{20}$,$(21, 6)^{21}$,$(22, 4)^{22}$,$(23, 2)^{23}$,$(24, 0)^{24}$,
$(21, 3)^{25}$,$(18, 6)^{26}$,$(15, 9)^{27}$,$(12,12)^{28}$,\\
$(10,13)^{29}$,$( 8,14)^{30}$,$( 6,15)^{31}$,$( 4,16)^{32}$,$( 3,15)^{33}$,
$( 2,14)^{34}$,$( 1,13)^{35}$,$( 0,12)^{36}$,$( 0, 9)^{37}$,\\
$( 0, 6)^{38}$,$( 0, 3)^{39}$,$( 0, 0)^{40}$ \\
$\eta=4$ \\
$( 8, 0)^{ 2}$,$(12, 0)^{ 3}$,$(16, 0)^{ 4}$,$(18, 1)^{ 5}$,$(20, 2)^{ 6}$,
$(22, 3)^{ 7}$,$(24, 4)^{ 8}$,$(27, 3)^{ 9}$,$(30, 2)^{10}$,\\
$(33, 1)^{11}$,$(36, 0)^{12}$,$(36, 2)^{13}$,$(36, 4)^{14}$,$(36, 6)^{15}$,
$(36, 8)^{16}$,$(37, 8)^{17}$,$(38, 8)^{18}$,$(39, 8)^{19}$,\\
$(40, 8)^{20}$,$(42, 6)^{21}$,$(44, 4)^{22}$,$(46, 2)^{23}$,$(48, 0)^{24}$,
$(46, 3)^{25}$,$(44, 6)^{26}$,$(42, 9)^{27}$,$(40,12)^{28}$,\\
$(39,13)^{29}$,$(38,14)^{30}$,$(37,15)^{31}$,$(36,16)^{32}$,$(36,15)^{33}$,
$(36,14)^{34}$,$(36,13)^{35}$,$(36,12)^{36}$,$(37, 9)^{37}$,\\
$(38, 6)^{38}$,$(39, 3)^{39}$,$(40, 0)^{40}$,$(36, 4)^{41}$,$(32, 8)^{42}$,
$(28,12)^{43}$,$(24,16)^{44}$,$(21,18)^{45}$,$(18,20)^{46}$,\\
$(15,22)^{47}$,$(12,24)^{48}$,$(10,24)^{49}$,$( 8,24)^{50}$,$( 6,24)^{51}$,
$( 4,24)^{52}$,$( 3,22)^{53}$,$( 2,20)^{54}$,$( 1,18)^{55}$,\\
$( 0,16)^{56}$,$( 0,12)^{57}$,$( 0, 8)^{58}$,$( 0, 4)^{59}$,$( 0, 0)^{60}$\\
\hline
\end{tabular}
\end{center}
\end{table}
\begin{table}
\begin{center}
\caption{Leading $SU(3)$ irrep $(\lambda_H , \mu_H)$ for a given number $m$ of 
nucleons and isospin $T=|T_Z|$. Results are  shown for $\eta = 2$, 3 and 
4 with $m$ even and $m \geq 4$. The irreps are shown the table as 
$(\lambda_H ,\mu_H)^{m,T}$.}
{\tiny{
\begin{tabular}{l}
\hline
$\eta=2$ \\
$( 8, 0)^{ 4, 0}$,$( 6, 1)^{ 4, 1}$,$( 4, 2)^{ 4, 2}$,$( 8, 2)^{ 6, 0}$,
$( 8, 2)^{ 6, 1}$,$( 7, 1)^{ 6, 2}$,$( 6, 0)^{ 6, 3}$,$( 8, 4)^{ 8, 0}$,
$( 9, 2)^{ 8, 1}$,\\
$(10, 0)^{ 8, 2}$,$( 6, 2)^{ 8, 3}$,$( 2, 4)^{ 8, 4}$,
$(10, 2)^{10, 0}$,$(10, 2)^{10, 1}$,$( 8, 3)^{10, 2}$,$( 6, 4)^{10, 3}$,
$( 3, 4)^{10, 4}$,$( 0, 4)^{10, 5}$,\\
$(12, 0)^{12, 0}$,$( 9, 3)^{12, 1}$,$( 6, 6)^{12, 2}$,$( 5, 5)^{12, 3}$,
$( 4, 4)^{12, 4}$,$( 2, 2)^{12, 5}$,
$( 0, 0)^{12, 6}$,$( 8, 4)^{14, 0}$,$( 8, 4)^{14, 1}$,$( 6, 5)^{14, 2}$,\\
$( 4, 6)^{14, 3}$,$( 4, 3)^{14, 4}$,
$( 4, 0)^{14, 5}$,$( 4, 8)^{16, 0}$,$( 5, 6)^{16, 1}$,$( 6, 4)^{16, 2}$,
$( 5, 3)^{16, 3}$,$( 4, 2)^{16, 4}$,\\
$( 2, 8)^{18, 0}$,$( 2, 8)^{18, 1}$,$( 4, 4)^{18, 2}$,$( 6, 0)^{18, 3}$,
$( 0, 8)^{20, 0}$,$( 1, 6)^{20, 1}$,
$( 2, 4)^{20, 2}$,$( 0, 4)^{22, 0}$,$( 0, 4)^{22, 1}$,$( 0, 0)^{24, 0}$\\
$\eta=3$ \\
$(12, 0)^{ 4, 0}$,$(10, 1)^{ 4, 1}$,$( 8, 2)^{ 4, 2}$,$(14, 2)^{ 6, 0}$,
$(14,2)^{ 6, 1}$,$(13, 1)^{ 6, 2}$,
$(12, 0)^{ 6, 3}$,$(16, 4)^{ 8, 0}$,$(17, 2)^{ 8, 1}$,\\
$(18, 0)^{ 8, 2}$,$(14, 2)^{ 8, 3}$,$(10, 4)^{ 8, 4}$,
$(20, 2)^{10, 0}$,$(20, 2)^{10, 1}$,$(18, 3)^{10, 2}$,$(16, 4)^{10, 3}$,
$(13, 4)^{10, 4}$,$(10, 4)^{10, 5}$,\\
$(24, 0)^{12, 0}$,$(21, 3)^{12, 1}$,$(18, 6)^{12, 2}$,$(17, 5)^{12, 3}$,
$(16, 4)^{12, 4}$,$(14, 2)^{12, 5}$,
$(12, 0)^{12, 6}$,$(22, 4)^{14, 0}$,$(22, 4)^{14, 1}$,\\
$(20, 5)^{14, 2}$,$(18, 6)^{14, 3}$,$(18, 3)^{14, 4}$,
$(18, 0)^{14, 5}$,$(12, 3)^{14, 6}$,$( 6, 6)^{14, 7}$,$(20, 8)^{16, 0}$,
$(21, 6)^{16, 1}$,$(22, 4)^{16, 2}$,\\
$(21, 3)^{16, 3}$,$(20, 2)^{16, 4}$,$(16, 4)^{16, 5}$,$(12, 6)^{16, 6}$,
$( 7, 7)^{16, 7}$,$( 2, 8)^{16, 8}$,
$(20, 8)^{18, 0}$,$(20, 8)^{18, 1}$,$(22, 4)^{18, 2}$,\\
$(24, 0)^{18, 3}$, $(19, 4)^{18, 4}$,$(14, 8)^{18, 5}$,$(11, 8)^{18, 6}$,
$( 8, 8)^{18, 7}$,$( 4, 7)^{18, 8}$,$( 0, 6)^{18, 9}$,
$(20, 8)^{20, 0}$,$(21, 6)^{20, 1}$,\\
$(22, 4)^{20, 2}$,$(20, 5)^{20, 3}$,$(18, 6)^{20, 4}$,$(14, 8)^{20, 5}$,
$(10,10)^{20, 6}$,$( 8, 8)^{20, 7}$,
$( 6, 6)^{20, 8}$,$( 3, 3)^{20, 9}$,$( 0, 0)^{20,10}$,\\
$(22, 4)^{22, 0}$,$(22, 4)^{22, 1}$,$(19, 7)^{22, 2}$,
$(16,10)^{22, 3}$,$(15, 9)^{22, 4}$,$(14, 8)^{22, 5}$,$(11, 8)^{22, 6}$,
$( 8, 8)^{22, 7}$,$( 7, 4)^{22, 8}$,\\
$( 6, 0)^{22, 9}$,$(24, 0)^{24, 0}$,$(20, 5)^{24, 1}$,$(16,10)^{24, 2}$,
$(14,11)^{24, 3}$,$(12,12)^{24, 4}$,
$(12, 9)^{24, 5}$,$(12, 6)^{24, 6}$,$(10, 4)^{24, 7}$,\\
$( 8, 2)^{24, 8}$,$(18, 6)^{26, 0}$,$(18, 6)^{26, 1}$,
$(15, 9)^{26, 2}$,$(12,12)^{26, 3}$,$(11,11)^{26, 4}$,$(10,10)^{26, 5}$,
$(11, 5)^{26, 6}$,$(12, 0)^{26, 7}$,\\
$(12,12)^{28, 0}$,$(13,10)^{28, 1}$,$(14, 8)^{28, 2}$,$(12, 9)^{28, 3}$,
$(10,10)^{28, 4}$,$(10, 7)^{28, 5}$,
$(10, 4)^{28, 6}$,$( 8,14)^{30, 0}$,$( 8,14)^{30, 1}$,\\
$(10,10)^{30, 2}$,$(12, 6)^{30, 3}$,$(11, 5)^{30, 4}$,
$(10, 4)^{30, 5}$,$( 4,16)^{32, 0}$,$( 5,14)^{32, 1}$,$( 6,12)^{32, 2}$,
$( 9, 6)^{32, 3}$,$(12, 0)^{32, 4}$,\\
$( 2,14)^{34, 0}$,$( 2,14)^{34, 1}$,$( 4,10)^{34, 2}$,$( 6, 6)^{34, 3}$,
$( 0,12)^{36, 0}$,$( 1,10)^{36, 1}$,
$( 2, 8)^{36, 2}$,$( 0, 6)^{38, 0}$,$( 0, 6)^{38, 1}$,$( 0, 0)^{40, 0}$\\
$\eta=4$ \\
$(16, 0)^{ 4, 0}$,$(14, 1)^{ 4, 1}$,$(12, 2)^{ 4, 2}$,$(20, 2)^{ 6, 0}$,
$(20, 2)^{ 6, 1}$,$(19, 1)^{ 6, 2}$,
$(18, 0)^{ 6, 3}$,$(24, 4)^{ 8, 0}$,$(25, 2)^{ 8, 1}$,\\
$(26, 0)^{ 8, 2}$,$(22, 2)^{ 8, 3}$,$(18, 4)^{ 8, 4}$,
$(30, 2)^{10, 0}$,$(30, 2)^{10, 1}$,$(28, 3)^{10, 2}$,$(26, 4)^{10, 3}$,
$(23, 4)^{10, 4}$,$(20, 4)^{10, 5}$,\\
$(36, 0)^{12, 0}$,$(33, 3)^{12, 1}$,$(30, 6)^{12, 2}$,$(29, 5)^{12, 3}$,
$(28, 4)^{12, 4}$,$(26, 2)^{12, 5}$,
$(24, 0)^{12, 6}$,$(36, 4)^{14, 0}$,$(36, 4)^{14, 1}$,\\
$(34, 5)^{14, 2}$,$(32, 6)^{14, 3}$,$(32, 3)^{14, 4}$,
$(32, 0)^{14, 5}$,$(26, 3)^{14, 6}$,$(20, 6)^{14, 7}$,$(36, 8)^{16, 0}$,
$(37, 6)^{16, 1}$,$(38, 4)^{16, 2}$,\\
$(37, 3)^{16, 3}$,$(36, 2)^{16, 4}$,$(32, 4)^{16, 5}$,$(28, 6)^{16, 6}$,
$(23, 7)^{16, 7}$,$(18, 8)^{16, 8}$,
$(38, 8)^{18, 0}$,$(38, 8)^{18, 1}$,$(40, 4)^{18, 2}$,\\
$(42, 0)^{18, 3}$,$(37, 4)^{18, 4}$,$(32, 8)^{18, 5}$,
$(29, 8)^{18, 6}$,$(26, 8)^{18, 7}$,$(22, 7)^{18, 8}$,$(18, 6)^{18, 9}$,
$(40, 8)^{20, 0}$,$(41, 6)^{20, 1}$,\\
$(42, 4)^{20, 2}$,$(40, 5)^{20, 3}$,$(38, 6)^{20, 4}$,$(34, 8)^{20, 5}$,
$(30,10)^{20, 6}$,$(28, 8)^{20, 7}$,
$(26, 6)^{20, 8}$,$(23, 3)^{20, 9}$,$(20, 0)^{20,10}$,\\
$(44, 4)^{22, 0}$,$(44, 4)^{22, 1}$,$(41, 7)^{22, 2}$,
$(38,10)^{22, 3}$,$(37, 9)^{22, 4}$,$(36, 8)^{22, 5}$,$(33, 8)^{22, 6}$,
$(30, 8)^{22, 7}$,$(29, 4)^{22, 8}$,\\
$(28, 0)^{22, 9}$,$(20, 4)^{22,10}$,$(12, 8)^{22,11}$,$(48, 0)^{24, 0}$,
$(44, 5)^{24, 1}$,$(40,10)^{24, 2}$,
$(38,11)^{24, 3}$,$(36,12)^{24, 4}$,$(36, 9)^{24, 5}$,\\
$(36, 6)^{24, 6}$,$(34, 4)^{24, 7}$,$(32, 2)^{24, 8}$,
$(26, 5)^{24, 9}$,$(20, 8)^{24,10}$,$(13,10)^{24,11}$,$( 6,12)^{24,12}$,
$(44, 6)^{26, 0}$,$(44, 6)^{26, 1}$,\\
$(41, 9)^{26, 2}$,$(38,12)^{26, 3}$,$(37,11)^{26, 4}$,$(36,10)^{26, 5}$,
$(37, 5)^{26, 6}$,$(38, 0)^{26, 7}$,
$(31, 5)^{26, 8}$,$(24,10)^{26, 9}$,$(19,11)^{26,10}$,\\
$(14,12)^{26,11}$,$( 8,12)^{26,12}$,$( 2,12)^{26,13}$,
$(40,12)^{28, 0}$,$(41,10)^{28, 1}$,$(42, 8)^{28, 2}$,$(40, 9)^{28, 3}$,
$(38,10)^{28, 4}$,$(38, 7)^{28, 5}$,\\
$(38, 4)^{28, 6}$,$(34, 6)^{28, 7}$,$(30, 8)^{28, 8}$,$(24,11)^{28, 9}$,
$(18,14)^{28,10}$,$(14,13)^{28,11}$,
$(10,12)^{28,12}$,$( 5,10)^{28,13}$,$( 0, 8)^{28,14}$,\\
$(38,14)^{30, 0}$,$(38,14)^{30, 1}$,$(40,10)^{30, 2}$,
$(42, 6)^{30, 3}$,$(41, 5)^{30, 4}$,$(40, 4)^{30, 5}$,$(35, 8)^{30, 6}$,
$(30,12)^{30, 7}$,$(27,12)^{30, 8}$,\\
$(24,12)^{30, 9}$,$(19,13)^{30,10}$,$(14,14)^{30,11}$,$(11,11)^{30,12}$,
$( 8, 8)^{30,13}$,$( 4, 4)^{30,14}$,
$( 0, 0)^{30,15}$,$(36,16)^{32, 0}$,$(37,14)^{32, 1}$,\\
$(38,12)^{32, 2}$,$(41, 6)^{32, 3}$,$(44, 0)^{32, 4}$,
$(38, 6)^{32, 5}$,$(32,12)^{32, 6}$,$(28,14)^{32, 7}$,$(24,16)^{32, 8}$,
$(22,14)^{32, 9}$,$(20,12)^{32,10}$,\\
$(16,11)^{32,11}$,$(12,10)^{32,12}$,$(10, 5)^{32,13}$,$( 8, 0)^{32,14}$,
$(36,14)^{34, 0}$,$(36,14)^{34, 1}$,
$(38,10)^{34, 2}$,$(40, 6)^{34, 3}$,$(38, 7)^{34, 4}$,\\
$(36, 8)^{34, 5}$,$(31,12)^{34, 6}$,$(26,16)^{34, 7}$,
$(23,16)^{34, 8}$,$(20,16)^{34, 9}$,$(19,12)^{34,10}$,$(18, 8)^{34,11}$,
$(15, 5)^{34,12}$,$(12, 2)^{34,13}$,\\
$(36,12)^{36, 0}$,$(37,10)^{36, 1}$,$(38, 8)^{36, 2}$,$(35,11)^{36, 3}$,
$(32,14)^{36, 4}$,$(31,13)^{36, 5}$,
$(30,12)^{36, 6}$,$(26,14)^{36, 7}$,$(22,16)^{36, 8}$,\\
$(20,14)^{36, 9}$,$(18,12)^{36,10}$,$(18, 6)^{36,11}$,
$(18, 0)^{36,12}$,$(38, 6)^{38, 0}$,$(38, 6)^{38, 1}$,$(34,11)^{38, 2}$,
$(30,16)^{38, 3}$,$(28,17)^{38, 4}$,\\
$(26,18)^{38, 5}$,$(26,15)^{38, 6}$,$(26,12)^{38, 7}$,$(23,12)^{38, 8}$,
$(20,12)^{38, 9}$,$(19, 8)^{38,10}$,
$(18, 4)^{38,11}$,$(40, 0)^{40, 0}$,$(35, 7)^{40, 1}$,\\
$(30,14)^{40, 2}$,$(27,17)^{40, 3}$,$(24,20)^{40, 4}$,
$(23,19)^{40, 5}$,$(22,18)^{40, 6}$,$(23,13)^{40, 7}$,$(24, 8)^{40, 8}$,
$(22, 6)^{40, 9}$,$(20, 4)^{40,10}$,\\
$(32, 8)^{42, 0}$,$(32, 8)^{42, 1}$,$(28,13)^{42, 2}$,$(24,18)^{42, 3}$,
$(22,19)^{42, 4}$,$(20,20)^{42, 5}$,
$(20,17)^{42, 6}$,$(20,14)^{42, 7}$,$(22, 7)^{42, 8}$,\\
$(24, 0)^{42, 9}$,$(24,16)^{44, 0}$,$(25,14)^{44, 1}$,
$(26,12)^{44, 2}$,$(23,15)^{44, 3}$,$(20,18)^{44, 4}$,$(19,17)^{44, 5}$,
$(18,16)^{44, 6}$,$(19,11)^{44, 7}$,\\
$(20, 6)^{44, 8}$,$(18,20)^{46, 0}$,$(18,20)^{46, 1}$,$(20,16)^{46, 2}$,
$(22,12)^{46, 3}$,$(20,13)^{46, 4}$,
$(18,14)^{46, 5}$,$(18,11)^{46, 6}$,$(18, 8)^{46, 7}$,\\
$(12,24)^{48, 0}$,$(13,22)^{48, 1}$,$(14,20)^{48, 2}$,
$(17,14)^{48, 3}$,$(20, 8)^{48, 4}$,$(19, 7)^{48, 5}$,$(18, 6)^{48, 6}$,
$( 8,24)^{50, 0}$,$( 8,24)^{50, 1}$,\\
$(10,20)^{50, 2}$,$(12,16)^{50, 3}$,$(16, 8)^{50, 4}$,$(20, 0)^{50, 5}$,
$( 4,24)^{52, 0}$,$( 5,22)^{52, 1}$,
$( 6,20)^{52, 2}$,$( 9,14)^{52, 3}$,$(12, 8)^{52, 4}$,\\
$( 2,20)^{54, 0}$,$( 2,20)^{54, 1}$,$( 4,16)^{54, 2}$,
$( 6,12)^{54, 3}$,$( 0,16)^{56, 0}$,$( 1,14)^{56, 1}$,$( 2,12)^{56, 2}$,
$( 0, 8)^{58, 0}$,$( 0, 8)^{58, 1}$,$( 0, 0)^{60, 0}$ \\
\hline
\end{tabular}
}}
\end{center}
\end{table}
\begin{table}
\begin{center}
\caption{Same as Table III but for $m$ odd. The leading $SU(3)$ irreps are
given in the table as $(\lambda_H ,\mu_H)^{m,2T}$.}
{\tiny{
\begin{tabular}{l}
\hline
$\eta=2$ \\
$( 6, 0)^{ 3, 1}$, $( 4, 1)^{ 3, 3}$, $( 8, 1)^{ 5, 1}$, $( 6, 2)^{ 5, 3}$, $( 5, 1)^{ 5, 5}$, $( 8, 3)^{ 7, 1}$, $( 9, 1)^{ 7, 3}$, $( 8, 0)^{ 7, 5}$, \\
$( 4, 2)^{ 7, 7}$, $( 9, 3)^{ 9, 1}$, $(10, 1)^{ 9, 3}$, $( 8, 2)^{ 9, 5}$, $( 4, 4)^{ 9, 7}$, $( 1, 4)^{ 9, 9}$, $(11, 1)^{11, 1}$, $( 8, 4)^{11, 3}$, \\
$( 6, 5)^{11, 5}$, $( 5, 4)^{11, 7}$, $( 2, 4)^{11, 9}$, $( 0, 2)^{11,11}$, $(10, 2)^{13, 1}$, $( 7, 5)^{13, 3}$, $( 5, 6)^{13, 5}$, $( 4, 5)^{13, 7}$, \\
$( 4, 2)^{13, 9}$, $( 2, 0)^{13,11}$, $( 6, 6)^{15, 1}$, $( 7, 4)^{15, 3}$, $( 5, 5)^{15, 5}$, $( 4, 4)^{15, 7}$, $( 4, 1)^{15, 9}$, $( 3, 8)^{17, 1}$, \\
$( 4, 6)^{17, 3}$, $( 6, 2)^{17, 5}$, $( 5, 1)^{17, 7}$, $( 1, 8)^{19, 1}$, $( 2, 6)^{19, 3}$, $( 4, 2)^{19, 5}$, $( 0, 6)^{21, 1}$, $( 1, 4)^{21, 3}$, \\
$( 0, 2)^{23, 1}$ \\
$\eta=3$ \\
$( 9, 0)^{ 3, 1}$, $( 7, 1)^{ 3, 3}$, $(13, 1)^{ 5, 1}$, $(11, 2)^{ 5, 3}$, $(10, 1)^{ 5, 5}$, $(15, 3)^{ 7, 1}$, $(16, 1)^{ 7, 3}$, $(15, 0)^{ 7, 5}$, \\
$(11, 2)^{ 7, 7}$, $(18, 3)^{ 9, 1}$, $(19, 1)^{ 9, 3}$, $(17, 2)^{ 9, 5}$, $(13, 4)^{ 9, 7}$, $(10, 4)^{ 9, 9}$, $(22, 1)^{11, 1}$, $(19, 4)^{11, 3}$, \\
$(17, 5)^{11, 5}$, $(16, 4)^{11, 7}$, $(13, 4)^{11, 9}$, $(11, 2)^{11,11}$, $(23, 2)^{13, 1}$, $(20, 5)^{13, 3}$, $(18, 6)^{13, 5}$, $(17, 5)^{13, 7}$, \\
$(17, 2)^{13, 9}$, $(15, 0)^{13,11}$, $( 9, 3)^{13,13}$, $(21, 6)^{15, 1}$, $(22, 4)^{15, 3}$, $(20, 5)^{15, 5}$, $(19, 4)^{15, 7}$, $(19, 1)^{15, 9}$, \\
$(15, 3)^{15,11}$, $( 9, 6)^{15,13}$, $( 4, 7)^{15,15}$, $(20, 8)^{17, 1}$, $(21, 6)^{17, 3}$, $(23, 2)^{17, 5}$, $(22, 1)^{17, 7}$, $(17, 5)^{17, 9}$, \\
$(13, 7)^{17,11}$, $(10, 7)^{17,13}$, $( 5, 8)^{17,15}$, $( 1, 7)^{17,17}$, $(20, 8)^{19, 1}$, $(21, 6)^{19, 3}$, $(23, 2)^{19, 5}$, $(21, 3)^{19, 7}$, \\
$(16, 7)^{19, 9}$, $(12, 9)^{19,11}$, $( 9, 9)^{19,13}$, $( 7, 7)^{19,15}$, $( 3, 6)^{19,17}$, $( 0, 3)^{19,19}$, $(21, 6)^{21, 1}$, $(22, 4)^{21, 3}$, \\
$(19, 7)^{21, 5}$, $(17, 8)^{21, 7}$, $(16, 7)^{21, 9}$, $(12, 9)^{21,11}$, $( 9, 9)^{21,13}$, $( 7, 7)^{21,15}$, $( 6, 3)^{21,17}$, $( 3, 0)^{21,19}$, \\
$(23, 2)^{23, 1}$, $(19, 7)^{23, 3}$, $(16,10)^{23, 5}$, $(14,11)^{23, 7}$, $(13,10)^{23, 9}$, $(13, 7)^{23,11}$, $(10, 7)^{23,13}$, $( 8, 5)^{23,15}$, \\ 
$( 7, 1)^{23,17}$, $(21, 3)^{25, 1}$, $(17, 8)^{25, 3}$, $(14,11)^{25, 5}$, $(12,12)^{25, 7}$, $(11,11)^{25, 9}$, $(11, 8)^{25,11}$, $(12, 3)^{25,13}$, \\
$(10, 1)^{25,15}$, $(15, 9)^{27, 1}$, $(16, 7)^{27, 3}$, $(13,10)^{27, 5}$, $(11,11)^{27, 7}$, $(10,10)^{27, 9}$, $(10, 7)^{27,11}$, $(11, 2)^{27,13}$, \\
$(10,13)^{29, 1}$, $(11,11)^{29, 3}$, $(13, 7)^{29, 5}$, $(11, 8)^{29, 7}$, $(10, 7)^{29, 9}$, $(10, 4)^{29,11}$, $( 6,15)^{31, 1}$, $( 7,13)^{31, 3}$, \\
$( 9, 9)^{31, 5}$, $(12, 3)^{31, 7}$, $(11, 2)^{31, 9}$, $( 3,15)^{33, 1}$, $( 4,13)^{33, 3}$, $( 6, 9)^{33, 5}$, $( 9, 3)^{33, 7}$, $( 1,13)^{35, 1}$, \\
$( 2,11)^{35, 3}$, $( 4, 7)^{35, 5}$, $( 0, 9)^{37, 1}$, $( 1, 7)^{37, 3}$, $( 0, 3)^{39, 1}$ \\
$\eta=4$ \\
$(12, 0)^{ 3, 1}$, $(10, 1)^{ 3, 3}$, $(18, 1)^{ 5, 1}$, $(16, 2)^{ 5, 3}$, $(15, 1)^{ 5, 5}$, $(22, 3)^{ 7, 1}$, $(23, 1)^{ 7, 3}$, $(22, 0)^{ 7, 5}$, \\
$(18, 2)^{ 7, 7}$, $(27, 3)^{ 9, 1}$, $(28, 1)^{ 9, 3}$, $(26, 2)^{ 9, 5}$, $(22, 4)^{ 9, 7}$, $(19, 4)^{ 9, 9}$, $(33, 1)^{11, 1}$, $(30, 4)^{11, 3}$, \\
$(28, 5)^{11, 5}$, $(27, 4)^{11, 7}$, $(24, 4)^{11, 9}$, $(22, 2)^{11,11}$, $(36, 2)^{13, 1}$, $(33, 5)^{13, 3}$, $(31, 6)^{13, 5}$, $(30, 5)^{13, 7}$, \\
$(30, 2)^{13, 9}$, $(28, 0)^{13,11}$, $(22, 3)^{13,13}$, $(36, 6)^{15, 1}$, $(37, 4)^{15, 3}$, $(35, 5)^{15, 5}$, $(34, 4)^{15, 7}$, $(34, 1)^{15, 9}$, \\
$(30, 3)^{15,11}$, $(24, 6)^{15,13}$, $(19, 7)^{15,15}$, $(37, 8)^{17, 1}$, $(38, 6)^{17, 3}$, $(40, 2)^{17, 5}$, $(39, 1)^{17, 7}$, $(34, 5)^{17, 9}$, \\
$(30, 7)^{17,11}$, $(27, 7)^{17,13}$, $(22, 8)^{17,15}$, $(18, 7)^{17,17}$, $(39, 8)^{19, 1}$, $(40, 6)^{19, 3}$, $(42, 2)^{19, 5}$, $(40, 3)^{19, 7}$, \\
$(35, 7)^{19, 9}$, $(31, 9)^{19,11}$, $(28, 9)^{19,13}$, $(26, 7)^{19,15}$, $(22, 6)^{19,17}$, $(19, 3)^{19,19}$, $(42, 6)^{21, 1}$, $(43, 4)^{21, 3}$, \\
$(40, 7)^{21, 5}$, $(38, 8)^{21, 7}$, $(37, 7)^{21, 9}$, $(33, 9)^{21,11}$, $(30, 9)^{21,13}$, $(28, 7)^{21,15}$, $(27, 3)^{21,17}$, $(24, 0)^{21,19}$, \\
$(16, 4)^{21,21}$, $(46, 2)^{23, 1}$, $(42, 7)^{23, 3}$, $(39,10)^{23, 5}$, $(37,11)^{23, 7}$, $(36,10)^{23, 9}$, $(36, 7)^{23,11}$, $(33, 7)^{23,13}$, \\ 
$(31, 5)^{23,15}$, $(30, 1)^{23,17}$, $(24, 4)^{23,19}$, $(16, 8)^{23,21}$, $( 9,10)^{23,23}$, $(46, 3)^{25, 1}$, $(42, 8)^{25, 3}$, $(39,11)^{25, 5}$, \\
$(37,12)^{25, 7}$, $(36,11)^{25, 9}$, $(36, 8)^{25,11}$, $(37, 3)^{25,13}$, $(35, 1)^{25,15}$, $(28, 6)^{25,17}$, $(22, 9)^{25,19}$, $(17,10)^{25,21}$, \\
$(10,12)^{25,23}$, $( 4,12)^{25,25}$, $(42, 9)^{27, 1}$, $(43, 7)^{27, 3}$, $(40,10)^{27, 5}$, $(38,11)^{27, 7}$, $(37,10)^{27, 9}$, $(37, 7)^{27,11}$, \\
$(38, 2)^{27,13}$, $(34, 4)^{27,15}$, $(27, 9)^{27,17}$, $(21,12)^{27,19}$, $(16,13)^{27,21}$, $(12,12)^{27,23}$, $( 6,12)^{27,25}$, $( 1,10)^{27,27}$, \\ 
$(39,13)^{29, 1}$, $(40,11)^{29, 3}$, $(42, 7)^{29, 5}$, $(40, 8)^{29, 7}$, $(39, 7)^{29, 9}$, $(39, 4)^{29,11}$, $(34, 8)^{29,13}$, $(30,10)^{29,15}$, \\
$(27,10)^{29,17}$, $(21,13)^{29,19}$, $(16,14)^{29,21}$, $(12,13)^{29,23}$, $( 9,10)^{29,25}$, $( 4, 8)^{29,27}$, $( 0, 4)^{29,29}$, $(37,15)^{31, 1}$, \\ 
$(38,13)^{31, 3}$, $(40, 9)^{31, 5}$, $(43, 3)^{31, 7}$, $(42, 2)^{31, 9}$, $(36, 8)^{31,11}$, $(31,12)^{31,13}$, $(27,14)^{31,15}$, $(24,14)^{31,17}$, \\
$(22,12)^{31,19}$, $(17,13)^{31,21}$, $(13,12)^{31,23}$, $(10, 9)^{31,25}$, $( 8, 4)^{31,27}$, $( 4, 0)^{31,29}$, $(36,15)^{33, 1}$, $(37,13)^{33, 3}$, \\
$(39, 9)^{33, 5}$, $(42, 3)^{33, 7}$, $(40, 4)^{33, 9}$, $(34,10)^{33,11}$, $(29,14)^{33,13}$, $(25,16)^{33,15}$, $(22,16)^{33,17}$, $(20,14)^{33,19}$, \\ 
$(19,10)^{33,21}$, $(15, 9)^{33,23}$, $(12, 6)^{33,25}$, $(10, 1)^{33,27}$, $(36,13)^{35, 1}$, $(37,11)^{35, 3}$, $(39, 7)^{35, 5}$, $(36,10)^{35, 7}$, \\
$(34,11)^{35, 9}$, $(33,10)^{35,11}$, $(28,14)^{35,13}$, $(24,16)^{35,15}$, $(21,16)^{35,17}$, $(19,14)^{35,19}$, $(18,10)^{35,21}$, $(18, 4)^{35,23}$, \\ 
$(15, 1)^{35,25}$, $(37, 9)^{37, 1}$, $(38, 7)^{37, 3}$, $(34,12)^{37, 5}$, $(31,15)^{37, 7}$, $(29,16)^{37, 9}$, $(28,15)^{37,11}$, $(28,12)^{37,13}$, \\
$(24,14)^{37,15}$, $(21,14)^{37,17}$, $(19,12)^{37,19}$, $(18, 8)^{37,21}$, $(18, 2)^{37,23}$, $(39, 3)^{39, 1}$, $(34,10)^{39, 3}$, $(30,15)^{39, 5}$, \\
$(27,18)^{39, 7}$, $(25,19)^{39, 9}$, $(24,18)^{39,11}$, $(24,15)^{39,13}$, $(25,10)^{39,15}$, $(22,10)^{39,17}$, $(20, 8)^{39,19}$, $(19, 4)^{39,21}$, \\ 
$(36, 4)^{41, 1}$, $(31,11)^{41, 3}$, $(27,16)^{41, 5}$, $(24,19)^{41, 7}$, $(22,20)^{41, 9}$, $(21,19)^{41,11}$, $(21,16)^{41,13}$, $(22,11)^{41,15}$, \\ 
$(24, 4)^{41,17}$, $(22, 2)^{41,19}$, $(28,12)^{43, 1}$, $(29,10)^{43, 3}$, $(25,15)^{43, 5}$, $(22,18)^{43, 7}$, $(20,19)^{43, 9}$, $(19,18)^{43,11}$, \\
$(19,15)^{43,13}$, $(20,10)^{43,15}$, $(22, 3)^{43,17}$, $(21,18)^{45, 1}$, $(22,16)^{45, 3}$, $(24,12)^{45, 5}$, $(21,15)^{45, 7}$, $(19,16)^{45, 9}$, \\
$(18,15)^{45,11}$, $(18,12)^{45,13}$, $(19, 7)^{45,15}$, $(15,22)^{47, 1}$, $(16,20)^{47, 3}$, $(18,16)^{47, 5}$, $(21,10)^{47, 7}$, $(19,11)^{47, 9}$, \\
$(18,10)^{47,11}$, $(18, 7)^{47,13}$, $(10,24)^{49, 1}$, $(11,22)^{49, 3}$, $(13,18)^{49, 5}$, $(16,12)^{49, 7}$, $(20, 4)^{49, 9}$, $(19, 3)^{49,11}$, \\
$( 6,24)^{51, 1}$, $( 7,22)^{51, 3}$, $( 9,18)^{51, 5}$, $(12,12)^{51, 7}$, $(16, 4)^{51, 9}$, $( 3,22)^{53, 1}$, $( 4,20)^{53, 3}$, $( 6,16)^{53, 5}$, \\ 
$( 9,10)^{53, 7}$, $( 1,18)^{55, 1}$, $( 2,16)^{55, 3}$, $( 4,12)^{55, 5}$, $( 0,12)^{57, 1}$, $( 1,10)^{57, 3}$, $( 0, 4)^{59, 1}$ \\
\hline
\end{tabular}
}}
\end{center}
\end{table}

\ed